\documentclass[10pt,twocolumn,aps,prd,nofootinbib]{revtex4-1}

\usepackage{amsmath}
\usepackage{amssymb}
\usepackage{graphicx}
\usepackage[utf8]{inputenc}
\usepackage[dvipsnames]{xcolor}

\usepackage[colorlinks,pdfusetitle]{hyperref}
\hypersetup{allcolors=[rgb]{1,0.56,0}}

\newcommand{\e}[1]{\times10^{#1}}

\begin{document}

\title{A Return To Neutrino Normalcy}

\author{Peter B.~Denton}

\date{\today}

\affiliation{High Energy Theory Group, Physics Department, Brookhaven National Laboratory, Upton, NY 11973, USA}
\email{pdenton@bnl.gov}

\begin{abstract}
Understanding the structure of the fermion mixing matrices is an important question in particle physics.
The quark mixing matrix is approximately diagonal while the lepton mixing matrix has large off-diagonal elements.
Attempting to understand these structures has been the focus of an large body of literature over the last several decades.
In this article we propose a new set of conditions to test the structure of mass matrices called normalcy based on how close to diagonal the mixing matrix is.
The mass ordering and the octant of $\theta_{23}$ represent two of these conditions.
We point out that the quark matrix easily satisfies all six normalcy conditions while none of them are known to be fully satisfied for leptons at high significance.
All of the conditions that can be tested for leptons suggest that the matrix could satisfy the normalcy conditions and upcoming experiments such as DUNE and T2HK will most likely determine if the lepton mass matrix satisfies all of them or not.
\end{abstract}

\maketitle

\section{Introduction}
The weak interaction acts in a different basis than the mass bases for both quarks and leptons \cite{Cabibbo:1963yz,Kobayashi:1973fv,Pontecorvo:1967fh,Maki:1962mu}.
While for quarks the two bases are very close together, for leptons the bases are quite different; understanding the structure of how these bases are related is an important open question in our understanding of the Standard Model (SM) of particle physics.
The quark mixing matrix is perturbative and is elegantly described by the Wolfenstein parameterization \cite{Wolfenstein:1983yz} which makes it quite clear not only that the matrix is approximately diagonal, but that as one moves away from the diagonal the elements fall off quite rapidly.
In contrast, the lepton mixing matrix includes much larger mixing and is clearly not perturbatively diagonal.
To this end there has been a large body of work to parameterize the lepton mixing matrix in the context of various symmetry groups, for some useful reviews see refs.~\cite{Altarelli:2010gt,King:2013eh,Xing:2015fdg}.
Alternatively, it has been suggested that the parameters in the lepton mixing matrix could be randomly drawn from a uniform distribution drawn over the Haar measure for a $3\times3$ complex unitary matrix; this suggestion goes by the name of anarchy \cite{Hall:1999sn,Haba:2000be,deGouvea:2003xe}\footnote{Models that posit that there is a combination of both flavor symmetries and some amount of anarchy also exist, see e.g.~\cite{Ge:2018ofp,Barrie:2019pqc}.}.
As upcoming neutrino oscillation experiments are expected to measure the lepton mixing matrix with a fair deal of precision, it is interesting to examine exactly what, if anything, can be determined about the fundamental nature of the lepton mixing matrix.
Symmetry approaches are difficult to either prove or disprove and, even in the presence of a perfect measurement, anarchy can only be disfavored at a given statistical significance.

In this article we propose a set of conditions that can be either verified or falsified limited only by the experimental precision of the measurement of the mixing parameters.
These conditions can provide insight into what is driving the large mixing in the lepton sector, which we dub normalcy conditions.
These conditions indicate if the neutrino mass eigenstates mix with the charged lepton states in the way we expect or not.
In most cases where normalcy can be tested given currently available data, the data prefers the normal case over the non-normal case.
Next generation oscillation experiments such as DUNE and T2HK \cite{Abi:2020evt,Abe:2014oxa} are necessary and should be sufficient to determine if all the normalcy conditions are simultaneously satisfied or not.

DUNE and JUNO are expected to determine the neutrino mass ordering in coming years \cite{Abi:2020evt,Djurcic:2015vqa}\footnote{Neutrino oscillation experiments are sensitive to two mass \emph{orderings}: normal and inverted, while absolute mass scale experiments are sensitive to three mass \emph{hierarchies}: normal, inverted, and quasi-degenerate.} and there is already a hint from global fits to oscillation data for the normal mass ordering at $\sim3\sigma$, although this has yet to be confirmed \cite{Capozzi:2017ipn,Esteban:2018azc,deSalas:2018bym,Capozzi:2018ubv,Kelly:2020fkv,Esteban:2020cvm}.
The mass ordering known as ``normal'' wherein $\Delta m^2_{31}>0$ provides the motivation for naming the remaining conditions the ``normalcy'' conditions.

Simply put, the normalcy conditions address the question of whether or not the lightest neutrino is mostly composed of the lightest charged lepton (the electron), least composed of the heaviest charged lepton (the tau), and all the permutations of this statement.

\section{Mass Eigenstate Definitions}
Before one can compare the relative size of different elements in the lepton mixing matrix, one needs to first make a careful definition of which mass eigenstate is labeled as $\nu_1$, $\nu_2$, and $\nu_3$.
Numerous definitions of the neutrino mass eigenstates exist in the literature.
In fact, the exact definition used in any given analysis is often not specified.
We will define our fiducial\footnote{Another commonly used definition, not listed here, treats the solar and atmospheric sectors differently, see the appendix.} mass eigenstate definition, labeled $\boldsymbol e$ to be
\begin{equation}
\boldsymbol e:\quad|U_{e1}|>|U_{e2}|>|U_{e3}|\,.
\label{eq:e}
\end{equation}
This definition has the advantage that we know from solar neutrinos \cite{Ahmad:2002jz} that there is one mass eigenstate that is $\sim2/3$ electron neutrino, another that is $\sim1/3$ electron neutrino, and the final one that is $\sim2\%$ electron neutrino from reactor neutrinos \cite{An:2012eh,Ahn:2012nd}.
Note that in this definition, $\theta_{12}<45^\circ$ by definition while the sign of $\Delta m^2_{21}$ needs to be measured\footnote{We use the standard PDG parameterization for the lepton mixing matrix in terms of three angles and one complex phase $U=O_{23}(\theta_{23})U_{13}(\theta_{13},\delta)O_{12}(\theta_{12})$ \cite{Tanabashi:2018oca}.}, however solar neutrino experiments measured the solar mass ordering to be normal at high significance: $\Delta m^2_{21}>0$ \cite{Ahmad:2002jz}.

Throughout this section we will compare other definitions to this one: if different definitions are equivalent then we say that the given normalcy condition is satisfied.

Another obvious choice assigns the mass eigenstates in increasing mass.
Labeled {\bf M} this definition is
\begin{equation}
{\bf M}:\quad m_1<m_2<m_3\,.
\end{equation}
This may well be the definition of choice in the future, especially if it is determined that definitions $\boldsymbol e$ and {\bf M} are equivalent.
The question of whether or not the {\bf M} condition is satisfied is the same as whether the atmospheric mass ordering is normal or inverted, hence the term ``normalcy''.

We note, however, that while definition $\boldsymbol e$ is preferential to electron neutrinos due to present experimental data, theoretically one should consider the corresponding definition for tau neutrinos equally,
\begin{equation}
\boldsymbol\tau:\quad|U_{\tau1}|<|U_{\tau2}|<|U_{\tau3}|\,.
\end{equation}
This is also perfectly valid definition, although given the small size of the global $\nu_\tau$ oscillation data set \cite{Agafonova:2018auq,Li:2017dbe,Aartsen:2019tjl}, it is not practical to take this as a fiducial definition.
Nonetheless this provides another normalcy condition; that is, we say that the lepton mixing matrix is normal if the set of $\boldsymbol\tau$ inequalities are satisfied when the mass eigenstates are defined as in eq.~\ref{eq:e}.

\section{Normalcy Conditions}
The $\boldsymbol e$ and $\boldsymbol\tau$ definitions are based on the rows (corresponding to unique flavors) of the mixing matrix.
One could imagine writing down similar definitions based on the columns (corresponding to unique masses),
\begin{align}
{\bf 1}:&\quad|U_{e1}|>|U_{\mu1}|>|U_{\tau1}|\,,\\
{\bf 3}:&\quad|U_{e3}|<|U_{\mu3}|<|U_{\tau3}|\,,
\end{align}
however there is no guarantee that either of these definitions uniquely define the three mass eigenstates.
Nonetheless, they do contribute two additional normalcy conditions: whether or not definition {\bf 1} ({\bf 3}) is satisfied simultaneously with $\boldsymbol e$ or not.

Two additional normalcy conditions exist relating the middle row and column of the matrix,
\begin{align}
{\bf 2}:&\quad|U_{\mu2}|>|U_{e2}|\text{ , and }|U_{\mu2}|>|U_{\tau2}|\,,\\
\boldsymbol\mu:&\quad|U_{\mu2}|>|U_{\mu1}|\text{ , and }|U_{\mu2}|>|U_{\mu3}|\,.\label{eq:mu}
\end{align}
These sets of inequalities differ from the others which all require all three numbers to be ordered, while these two only require that one number ($|U_{\mu2}|$) is larger than two other numbers.

\begin{figure}
\centering
\includegraphics[width=\columnwidth]{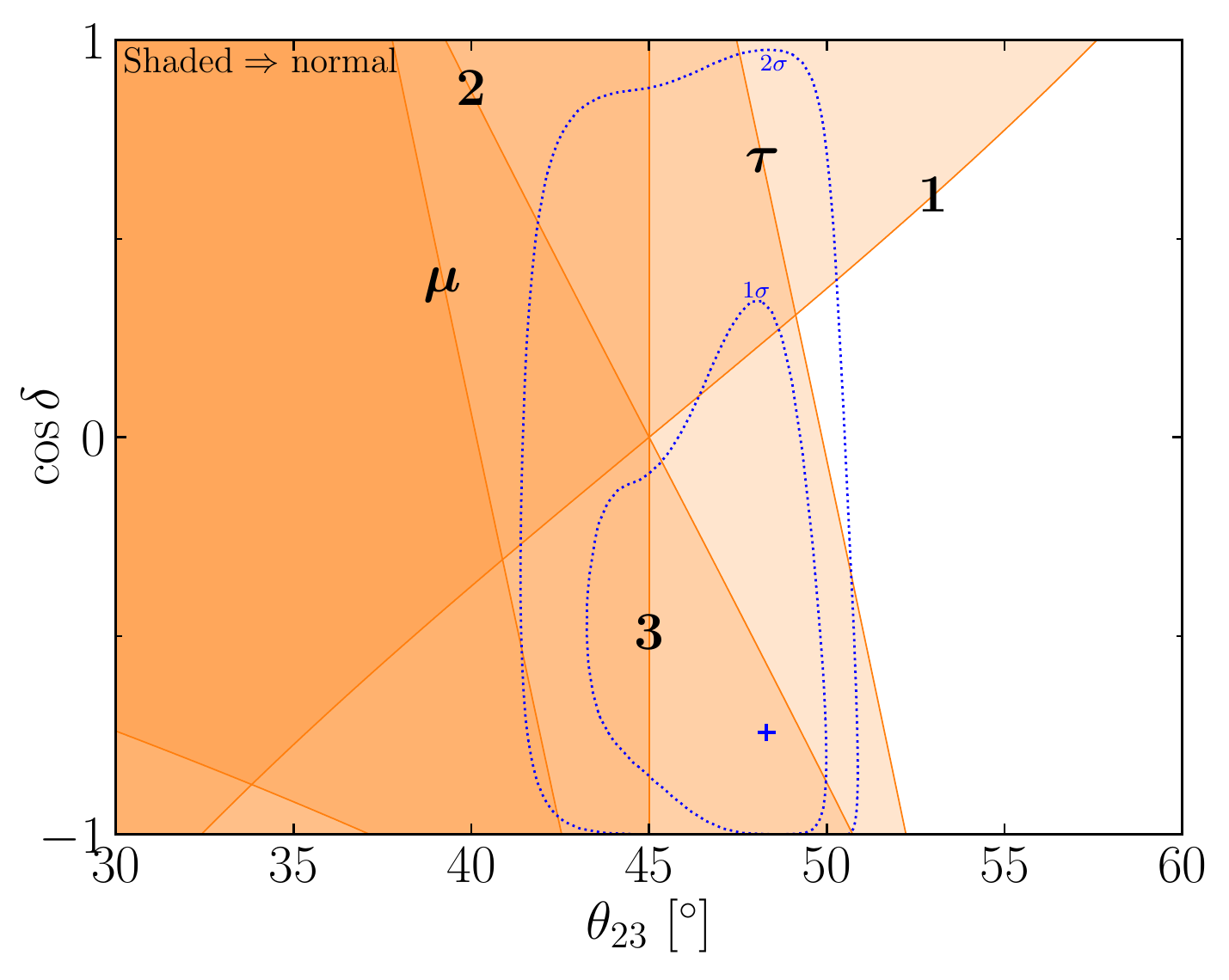}
\caption{The orange shaded regions to the left are the allowed normalcy regions assuming $\theta_{13}=8.61^\circ$ and $\theta_{12}=33.82^\circ$.
The best fit point and $\Delta\chi^2=2.3,6.18$ corresponding approximately to $1,2\sigma$ from the nufit 4.1 global fit (normal mass ordering and without Super-K atmospheric data) \cite{Esteban:2018azc} are shown as a blue plus and dotted contours respectively.
The light region in the lower-left corner is where the first inequality in the $\boldsymbol\tau$ condition no longer holds, although this region is quite disfavored by the data.}
\label{fig:conditions}
\end{figure}

Next, we examine when each of these conditions are satisfied in terms of the usual four mixing parameters: $\theta_{12}$, $\theta_{13}$, $\theta_{23}$, and $\delta$.
The second inequality in definition {\bf 3} is satisfied exactly when $\theta_{23}<45^\circ$; that is $\theta_{23}$ being in the first octant is a normalcy condition.
For the remaining conditions, the inequalities involving the $\mu$ and $\tau$ rows and 1 and 2 columns don't have simple exact solutions in the usual parameterization of the mixing matrix, but the second inequality (the one that hasn't been experimentally determined yet) in each case is well approximated by,
\begin{align}
{\bf 1}:\quad\cos\delta\gtrsim{}&-\frac{s_{12}\cos2\theta_{23}}{2c_{12}s_{13}}\simeq-2.2\cos2\theta_{23}\,,\\
{\bf 2}:\quad\cos\delta\lesssim{}&\frac{c_{12}\cos2\theta_{23}}{2s_{12}s_{13}}\simeq5.0\cos2\theta_{23}\,,\\
\boldsymbol\mu:\quad\cos\delta\lesssim{}&-\frac{(c_{13}^2+c_{12}^2-s_{12}^2s_{13}^2)s_{23}^2-c_{12}^2}{c_{12}s_{12}s_{13}}\nonumber\\
&\qquad\qquad\qquad\qquad\simeq10.0-24.0s_{23}^2\,,\\
\boldsymbol\tau:\quad\cos\delta\lesssim{}&-\frac{(c_{13}^2+c_{12}^2-s_{12}^2s_{13}^2)s_{23}^2-c_{13}^2+s_{12}^2s_{13}^2}{c_{12}s_{12}s_{13}}\nonumber\\
&\qquad\qquad\qquad\qquad\simeq14.0-24.0s_{23}^2\,,
\end{align}
where we used the fact that the doubly reduced Jarlskog \cite{Jarlskog:1985ht,Denton:2016wmg} $J_{rr}\equiv s_{12}c_{12}s_{23}c_{23}s_{13}\approx\frac12s_{12}c_{12}s_{13}$ for $\theta_{23}\sim45^\circ$.
The exact inequalities are numerically plotted in fig.~\ref{fig:conditions} assuming $\theta_{13}=8.61^\circ$ and $\theta_{12}=33.82^\circ$ \cite{Esteban:2018azc,peter_b_denton_2020_3709501}.

We note that conditions {\bf 1}, {\bf 2}, and {\bf 3} all cross at exactly $\theta_{23}=45^\circ$ and $\cos\delta=0$.
The $\theta_{23}=45^\circ$ statement results from the fact that the second inequality in condition {\bf 3} is exactly whether $\theta_{23}$ is less than $45^\circ$ or not.
Then we use the fact that $\mu$-$\tau$ symmetry is exactly satisfied when $\theta_{23}=45^\circ$ and $\cos\delta=0$, as can easily be seen by expanding the square of the norm of the elements,
\begin{align*}
|U_{\mu1}|^2&=s_{12}^2c_{23}^2+c_{12}^2s_{23}^2s_{13}^2+2s_{12}c_{12}s_{23}c_{23}s_{13}\cos\delta\,,\\
|U_{\tau1}|^2&=s_{12}^2s_{23}^2+c_{12}^2c_{23}^2s_{13}^2-2s_{12}c_{12}s_{23}c_{23}s_{13}\cos\delta\,,
\end{align*}
for the first column and,
\begin{align*}
|U_{\mu2}|^2&=c_{12}^2c_{23}^2+s_{12}^2s_{23}^2s_{13}^2-2s_{12}c_{12}s_{23}c_{23}s_{13}\cos\delta\,,\\
|U_{\tau2}|^2&=c_{12}^2s_{23}^2+s_{12}^2c_{23}^2s_{13}^2+2s_{12}c_{12}s_{23}c_{23}s_{13}\cos\delta\,,
\end{align*}
for the second column.
In either case the first two terms are equal when $\theta_{23}=45^\circ$ and the third terms are equal when $\cos\delta=0$.

While there are six different normalcy conditions, in most cases given the current oscillation picture, only two inequalities are required to be satisfied in order to satisfy all the normalcy conditions as can be seen in fig.~\ref{fig:conditions},
\begin{equation}
|U_{\mu2}|>|U_{\mu3}|\quad\text{and}\quad|U_{\mu1}|>|U_{\tau1}|\,,
\end{equation}
which can be approximated as,
\begin{align}
\cos\delta\gtrsim{}&4.4s_{23}^2-2.2\,,\label{eq:approxcds231}\\
\cos\delta\lesssim{}&10.0-24.0s_{23}^2\,.\label{eq:approxcds232}
\end{align}
We have numerically verified that all of these approximations are very accurate.

\begin{figure}
\centering
\includegraphics[width=0.7\columnwidth,bb=72 96 193 171,clip=true]{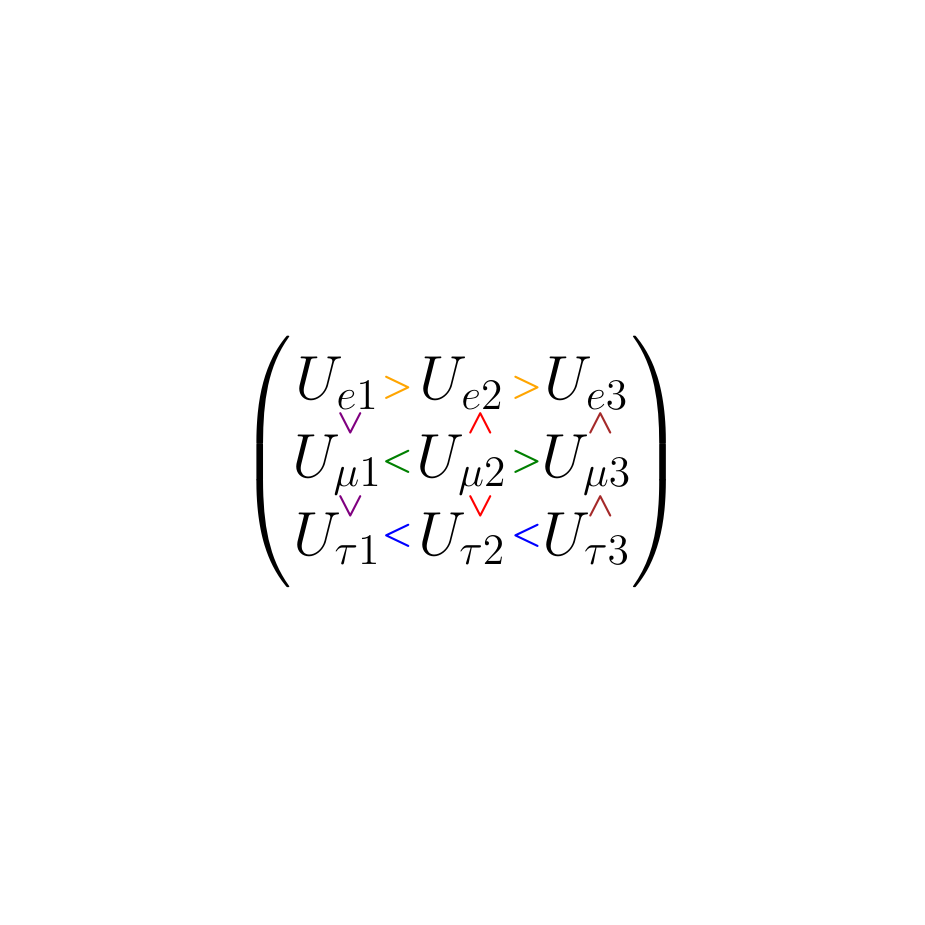}
\caption{The different inequalities among the absolute value of the elements of the mixing matrix are shown schematically.
The conditions are color coded: $\boldsymbol e$ in orange (the fiducial definition of the mass eigenstates), $\boldsymbol\mu$ in green, $\boldsymbol\tau$ in blue, {\bf 1} in purple, {\bf 2} in red, and {\bf 3} in brown.}
\label{fig:schematic}
\end{figure}

To summarize, assuming that the mass eigenstates are defined by definition $\boldsymbol e$ given in eq.~\ref{eq:e}, the six (convention independent) normalcy conditions are whether or not the following conditions are true:
\begin{align*}
{\bf M}:&\quad m_1<m_2<m_3\,,\\
\boldsymbol\mu:&\quad|U_{\mu2}|>|U_{\mu1}|\quad\text{and}\quad|U_{\mu2}|>|U_{\mu3}|\,,\\
\boldsymbol\tau:&\quad|U_{\tau1}|<|U_{\tau2}|<|U_{\tau3}|\,,\\
{\bf 1}:&\quad|U_{e1}|>|U_{\mu1}|>|U_{\tau1}|\,,\\
{\bf 2}:&\quad|U_{\mu2}|>|U_{e2}|\quad\text{and}\quad|U_{\mu2}|>|U_{\tau2}|\,,\\
{\bf 3}:&\quad|U_{e3}|<|U_{\mu3}|<|U_{\tau3}|\,.
\end{align*}
To a good approximation this means that if $\cos\delta$ and $s_{23}^2$ satisfy the relationships in eqs.~\ref{eq:approxcds231}-\ref{eq:approxcds232} and the mass ordering is normal then all the normalcy conditions are satisfied.
Schematically the six sets of inequalities relating to the lepton mixing matrix including the fiducial definition $\boldsymbol e$ can be seen in fig.~\ref{fig:schematic}.

\section{Current Status and Future Prospects}
Current global fits indicate that some parts of each condition are satisfied, but none of them are completely satisfied.
The $3\sigma$ allowed region for the absolute value of each element in the lepton mixing matrix is \cite{Esteban:2018azc}
\begin{equation*}
\begin{pmatrix}
0.797\to0.842&0.518\to0.585&0.143\to0.156\\
0.244\to0.496&0.467\to0.678&0.646\to0.772\\
0.287\to0.525&0.488\to0.693&0.618\to0.749
\end{pmatrix}\,.
\end{equation*}
Given that we know that $\Delta m^2_{21}>0$ from solar experiments, we know that the {\bf M} condition is partially true.
If the atmospheric mass ordering is confirmed to be normal then the {\bf M} condition is true.
For the $\tau$ row condition, $\boldsymbol\tau$, we see $|U_{\tau1}|<|U_{\tau3}|$, but $|U_{\tau2}|$ could be anywhere relative to the other two.

For the mass eigenstate conditions {\bf 3} and {\bf 1}, we see that the electron component of the conditions are known to be satisfied.
For condition {\bf 3} that means that we know that $|U_{e3}|<|U_{\mu3}|$ and $|U_{e3}|<|U_{\tau3}|$, but we don't know which of $|U_{\mu3}|$ and $|U_{\tau3}|$ are larger.
Similarly, for condition {\bf 1} that means that we know that $|U_{e1}|>|U_{\mu1}|$ and $|U_{e1}|>|U_{\tau1}|$, but we don't know which of $|U_{\mu1}|$ and $|U_{\tau1}|$ are larger.

The picture for the middle row and column is less clear as none of the inequalities are known to be true and some are disfavored at $\sim2\sigma$ ($\Delta\chi^2\sim6$ for the $\boldsymbol\mu$ condition).

The poorest measured oscillation parameters are $\theta_{23}$ and $\delta$ which govern essentially all of the remaining uncertainty in the normalcy conditions.
DUNE and T2HK both expect to measure $\theta_{23}$ with $\lesssim1^\circ$ resolution and $\delta$ should be measured with $\sim10-15^\circ$ precision from each experiment alone \cite{Abe:2014oxa,Abi:2020evt}.
It may be the case that some of these inequalities may be close enough together to make a determination on normalcy quite difficult.
In this case a high level of precision will be needed to determine which is larger than the other, although such a strong relationship among elements may be interpreted as an indication of a symmetry present in the lepton mixing matrix.
However, we expect that DUNE and T2HK, certainly with addition of other oscillation data in a global fit, should be able to determine if these normalcy conditions are satisfied or not.

We briefly comment on the quark sector.
The allowed values of the absolute value of the elements of the quark mixing matrix at $3\sigma$ are \cite{Tanabashi:2018oca}\footnote{The $3\sigma$ allowed values are taken to be three times the $1\sigma$ allowed regions. We also note that allowed values in the quark matrix are determined without an assumption of unitarity, while those in the lepton matrix do assume unitarity. The lepton numbers without unitarity can be found in ref.~\cite{Parke:2015goa} and are even less restrictive, although the {\bf e} definition is still well defined.}
\begin{equation*}
\begin{pmatrix}
0.974\to0.975&0.223\to0.226&0.003\to0.005\\
0.206\to0.230&0.946\to1.048&0.040\to0.045\\
0.007\to0.010&0.033\to0.046&0.944\to1.094
\end{pmatrix}\,,
\end{equation*}
from which we can see that each normalcy definition is clearly satisfied by also noting that the matrix is defined with both up-like and down-like quarks arranged with increasing mass.

\section{Flavor Symmetries}
While the notion of normalcy is not directly derived from a symmetry group, it has conceptual similarities to certain symmetry groups.
In this section we consider several other theoretically motivated predictions and means of quantifying the expected parameters of the lepton mixing matrix.

\subsection{Zero parameter ansatzes}
Before the measurement of non-zero $\theta_{13}$, numerous neutrino mixing ansatzes were presented to describe the structure of the first two mixing angles measured: $\theta_{23}$ and $\theta_{12}$.
While these are now in disagreement with the data, they provide an important jumping off point for more involved ansatzes.
In table \ref{tab:symmetries} we show which of the popular symmetries with zero free parameters satisfy the individual normalcy conditions.
Most of them including tribimaximal (TBM) mixing do not satisfy the $\boldsymbol\mu$ condition; only the trimaximal (TM) mixing satisfies all the normalcy conditions which requires allowing the inequalities to be non-strict: $<$ $\to$ $\le$.

\begin{table}
\newcommand{\y}{{\color{ForestGreen}\checkmark}}
\newcommand{\n}{{$\color{red}\times$}}
\centering
\caption{Here we show if various symmetries satisfy the individual and overall normalcy conditions.
The symmetries mentioned are
bimaximal (BM) \cite{Vissani:1997pa}, trimaximal (TM) \cite{Harrison:1994iv}, democratic (D) \cite{Harrison:2002er}, tribimaximal (TBM) \cite{Harrison:2002er}, and golden ratio (GR) \cite{Everett:2008et}.
Some of the conditions are only satisfied if the inequalities are relaxed to non-strict: $<$ $\to$ $\le$.}
\begin{tabular}{c|c|c|c|c|c|c|c}
&$\boldsymbol M$&$\boldsymbol\mu$&$\boldsymbol\tau$&\bf1&\bf2&\bf3&Normalcy\\\hline
BM&\y&\n&\y&\y&\n&\y&\n\\
TM&\y&\y&\y&\y&\y&\y&\y\\
D&\y&\n&\y&\y&\y&\y&\n\\
TBM&\y&\n&\y&\y&\y&\y&\n\\
GR&\y&\n&\y&\y&\y&\y&\n
\end{tabular}
\label{tab:symmetries}
\end{table}

\subsection{Generalized ansatzes}
\begin{figure}
\centering
\includegraphics[width=\columnwidth]{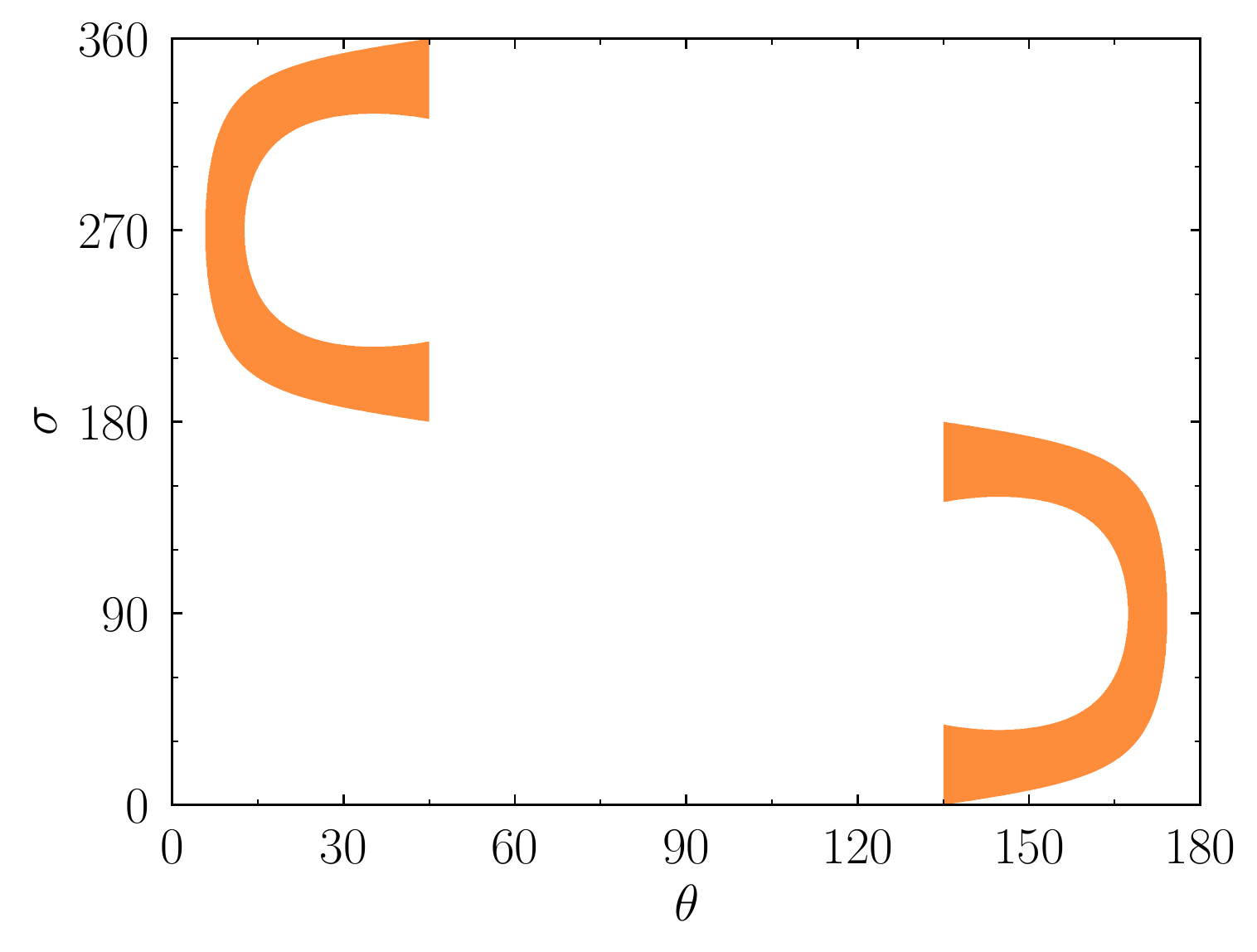}
\caption{The shaded regions of the parameters in the generalized tribimaximal mixing \cite{Chen:2018eou} satisfy all of the normalcy conditions; the allowed regions are independent of $\rho$.
Some of the conditions are only satisfied if elements are allowed to be equal to each other.}
\label{fig:gTBM}
\end{figure}

As most of these symmetries predict $\theta_{13}=0$ which is strongly ruled out, many of them have been extended in recognition of the data.
One such example is generalized tribimaximal (gTBM) mixing \cite{Chen:2018eou} where the mixing matrix is parameterized as a function of three parameters $\theta$, $\rho$, and $\sigma$,
\begin{multline}
\setlength\arraycolsep{4pt}
U_{\rm gTBM}=\\
\begin{pmatrix}
\sqrt{\frac23}&\frac{e^{-i\rho}\cos\theta}{\sqrt3}&-\frac{ie^{-i\rho}\sin\theta}{\sqrt3}\\[4pt]
-\frac{e^{i\rho}}{\sqrt6}&\frac{\cos\theta}{\sqrt3}-\frac{ie^{-i\sigma}\sin\theta}{\sqrt2}&-\frac{i\sin\theta}{\sqrt3}+\frac{e^{-i\sigma}\cos\theta}{\sqrt2}\\[4pt]
\frac{e^{i(\rho+\sigma)}}{\sqrt6}&-\frac{i\sin\theta}{\sqrt2}-\frac{e^{i\sigma}\cos\theta}{\sqrt3}&\frac{\cos\theta}{\sqrt2}+\frac{ie^{i\sigma}\sin\theta}{\sqrt3}
\end{pmatrix}\,.
\end{multline}
Note that as $\theta$, $\rho$, and $\sigma\to0$, TBM mixing is recovered.
Since we are only interested in the absolute value of the elements of the matrix, $\rho$ as no effect.
We find that gTBM mixing satisfies the normalcy conditions for certain ranges of $\theta$ and $\sigma$ (and all values of $\rho$) shown in fig.~\ref{fig:gTBM} provided that some of the normalcy inequalities are relaxed $<$ $\to$ $\le$.
If strict inequalities are enforced in the normalcy conditions, then there are no normalcy regions for gTBM.

\subsection{Neutrino anarchy}
Neutrino anarchy \cite{Haba:2000be} posits that the neutrino mixing parameters are likely to be uniformly sampled from a generic $3\times3$ matrix.
The correct metric for the sampling is given by the Haar measure and non-trivially depends on whether the matrix is real or not.
Since there is no reason to expect that CP is conserved (and there have already been some hints of CP violation in the leptonic mass matrix \cite{Abe:2018wpn}), we focus on the Haar measure for complex matrices,
\begin{equation}
d\Pi=ds_{12}^2dc_{13}^4ds_{23}^2d\delta\,,
\end{equation}
where the contribution from additional unphysical and Majorana (unphysical for oscillations) phases has been suppressed.
We now evaluate the fraction of parameter space that satisfies the normalcy conditions,
\begin{equation}
\frac{\int_{\rm normalcy}d\Pi}{\int d\Pi}=0.022\,.
\end{equation}
That is, normalcy is somewhat unlikely within the anarchy picture as only 2.2\% of the parameter space satisfies the normalcy conditions.

\section{Discussion}
These seven sets of inequalities (eqs.~\ref{eq:e}-\ref{eq:mu}) can be combined into six normalcy conditions, most of which are known to be partially satisfied in the lepton sector suggesting that normalcy may be a valid guiding principle for the lepton mass matrix.
While many symmetry groups predict $\theta_{23}=45^\circ$, this leads to $|U_{\mu2}|<|U_{\mu3}|$ which is non-normal, thus many zero-parameter flavor ansatzes actually predict a deviation from normalcy, see table \ref{tab:symmetries}.
We have shown that in more generalized ansatzes normalcy predicts a specific correlation among the parameters, one such example is shown in fig.~\ref{fig:gTBM}.

In the anarchy scenario, normalcy is not that likely; it is disfavored at the 2.2\% level which is $>2\sigma$.
While not at high confidence, a measurement of normalcy could be seen as putting some tension on the anarchy hypothesis.

Anarchy, while consistent with the lepton sector at the $p=0.41$ ($0.08$) level, it is disfavored in the quark sector at the $p=6\e{-6}$ ($10^{-4}$) level where the first (second) number referring to different choices of the prior \cite{deGouvea:2003xe}.
On the other hand, normalcy has the possibility to be satisfied for both quarks and leptons indicating that it may be the guiding rule for describing both mass mixing matrices.
While the normalcy criteria have been determined \textit{a posteriori} since they are already known to be partially satisfied in the lepton sector (and fully satisfied in the quark sector), since their determination is not statistical in nature an \textit{a priori} definition would look the same as the one described in this article.

This normalcy tool thus provides a guiding principle that can be easily interpreted and will mostly likely be either confirmed or ruled out by upcoming experiments.
In addition to the obvious importance that measuring CP violation in neutrino has, these normalcy conditions highlight the fact that the neutrino mass ordering and the $\theta_{23}$ octant are two parts in a larger set of normalcy conditions to be determined.
Whether or not the remaining normalcy conditions can be determined depends on the precision with which $\theta_{23}$ and $\cos\delta$ can be measured.

The best fit region given the global oscillation picture that satisfies all the normalcy conditions is found at $\theta_{23}\simeq41^\circ$ and $\cos\delta\simeq-0.31$ for $\theta_{12}$ and $\theta_{13}$ fixed (that is, $\delta\simeq252^\circ$ given current T2K hints for $\sin\delta\lesssim0$ \cite{Abe:2018wpn}).
This point is mildly disfavored from the best fit non-normal point at $\Delta\chi^2=8$ which is approximately $2.4\sigma$ assuming 2 dof's.

\section{Conclusion}
In this article we stated three definitions of neutrino mass eigenstate numbering.
One of them ({\bf M}) is based on masses of the neutrinos while the other two ($\boldsymbol e$ and $\boldsymbol\tau$) are based on the relative components of the electron and tau neutrinos.
Whether or not these are equivalent as one would expect provide two normalcy conditions.
Four additional normalcy conditions ({\bf 1}, {\bf 2}, {\bf 3}, and $\boldsymbol\mu$) are based on the relative components of $\nu_1$, $\nu_2$, $\nu_3$, and $\nu_\mu$ respectively\footnote{There are seven sets of inequalities (any one out of three of them can be used as definitions of the mass eigenstates), one of which ($\boldsymbol e$) is used as the fiducial definition of $\nu_1$, $\nu_2$, and $\nu_3$.
The remaining sets of inequalities provide six normalcy conditions.}.
Each of the six normalcy conditions is essentially requiring that the more off-diagonal elements are smaller than the more diagonal elements.
Put another way, normalcy requires that the lightest neutrino is mostly connected with the lightest charged lepton, and so on.

As normalcy is somewhat unlikely in the anarchy hypothesis, a detection of normalcy may put some slight strain on the anarchy hypothesis, although the significance in question is low.
In addition, while most zero-parameter flavor mixing ansatzes violate one or more normalcy conditions, generalized versions do allow for all the normalcy conditions to be simultaneously satisfied.
Thus normalcy  predicts a certain correlation among parameters in generalized models.

\begin{acknowledgments}
We thank Julia Gehrlein, Boris Kayser, and Stephen Parke for useful comments.
Work supported by the US Department of Energy under Grant Contract DE-SC0012704.
\end{acknowledgments}

\appendix
\section{Appendix: Solar Mass Eigenstate Definition}
\label{sec:solar}
In solar neutrino analyses $\Delta m^2_{21}>0$ is often taken to be a definition.
Then the solar mass ordering question is replaced as a question on the octant of the solar angle.
That is, whether or not $\theta_{12}<45^\circ$ (lower octant) or $\theta_{12}>45^\circ$.

Thus the definition of the mass eigenstates in these analyses is sometimes
\begin{equation}
m_1<m_2\quad\text{and}\quad|U_{e3}|<|U_{e1}|\quad\text{and}\quad|U_{e3}|<|U_{e2}|\,.
\end{equation}
We prefer to avoid this definition as it requires mixing two different kinds of definitions: one about the relative size of the terms in the electron neutrino row and one about the ordering of the masses.

\bibliography{Normalcy}

\begin{thebibliography}{39}%
\makeatletter
\providecommand \@ifxundefined [1]{%
 \@ifx{#1\undefined}
}%
\providecommand \@ifnum [1]{%
 \ifnum #1\expandafter \@firstoftwo
 \else \expandafter \@secondoftwo
 \fi
}%
\providecommand \@ifx [1]{%
 \ifx #1\expandafter \@firstoftwo
 \else \expandafter \@secondoftwo
 \fi
}%
\providecommand \natexlab [1]{#1}%
\providecommand \enquote  [1]{``#1''}%
\providecommand \bibnamefont  [1]{#1}%
\providecommand \bibfnamefont [1]{#1}%
\providecommand \citenamefont [1]{#1}%
\providecommand \href@noop [0]{\@secondoftwo}%
\providecommand \href [0]{\begingroup \@sanitize@url \@href}%
\providecommand \@href[1]{\@@startlink{#1}\@@href}%
\providecommand \@@href[1]{\endgroup#1\@@endlink}%
\providecommand \@sanitize@url [0]{\catcode `\\12\catcode `\$12\catcode
  `\&12\catcode `\#12\catcode `\^12\catcode `\_12\catcode `\%12\relax}%
\providecommand \@@startlink[1]{}%
\providecommand \@@endlink[0]{}%
\providecommand \url  [0]{\begingroup\@sanitize@url \@url }%
\providecommand \@url [1]{\endgroup\@href {#1}{\urlprefix }}%
\providecommand \urlprefix  [0]{URL }%
\providecommand \Eprint [0]{\href }%
\providecommand \doibase [0]{http://dx.doi.org/}%
\providecommand \selectlanguage [0]{\@gobble}%
\providecommand \bibinfo  [0]{\@secondoftwo}%
\providecommand \bibfield  [0]{\@secondoftwo}%
\providecommand \translation [1]{[#1]}%
\providecommand \BibitemOpen [0]{}%
\providecommand \bibitemStop [0]{}%
\providecommand \bibitemNoStop [0]{.\EOS\space}%
\providecommand \EOS [0]{\spacefactor3000\relax}%
\providecommand \BibitemShut  [1]{\csname bibitem#1\endcsname}%
\let\auto@bib@innerbib\@empty
\bibitem [{\citenamefont {Cabibbo}(1963)}]{Cabibbo:1963yz}%
  \BibitemOpen
  \bibfield  {author} {\bibinfo {author} {\bibfnamefont {N.}~\bibnamefont
  {Cabibbo}},\ }\bibfield  {booktitle} {\emph {\bibinfo {booktitle} {{Meeting
  of the Italian School of Physics and Weak Interactions Bologna, Italy, April
  26-28, 1984}}},\ }\href {\doibase 10.1103/PhysRevLett.10.531} {\bibfield
  {journal} {\bibinfo  {journal} {Phys. Rev. Lett.}\ }\textbf {\bibinfo
  {volume} {10}},\ \bibinfo {pages} {531} (\bibinfo {year} {1963})},\ \bibinfo
  {note} {[648(1963)]}\BibitemShut {NoStop}%
\bibitem [{\citenamefont {Kobayashi}\ and\ \citenamefont
  {Maskawa}(1973)}]{Kobayashi:1973fv}%
  \BibitemOpen
  \bibfield  {author} {\bibinfo {author} {\bibfnamefont {M.}~\bibnamefont
  {Kobayashi}}\ and\ \bibinfo {author} {\bibfnamefont {T.}~\bibnamefont
  {Maskawa}},\ }\href {\doibase 10.1143/PTP.49.652} {\bibfield  {journal}
  {\bibinfo  {journal} {Prog. Theor. Phys.}\ }\textbf {\bibinfo {volume}
  {49}},\ \bibinfo {pages} {652} (\bibinfo {year} {1973})}\BibitemShut
  {NoStop}%
\bibitem [{\citenamefont {Pontecorvo}(1968)}]{Pontecorvo:1967fh}%
  \BibitemOpen
  \bibfield  {author} {\bibinfo {author} {\bibfnamefont {B.}~\bibnamefont
  {Pontecorvo}},\ }\href@noop {} {\bibfield  {journal} {\bibinfo  {journal}
  {Sov. Phys. JETP}\ }\textbf {\bibinfo {volume} {26}},\ \bibinfo {pages} {984}
  (\bibinfo {year} {1968})},\ \bibinfo {note} {[Zh. Eksp. Teor.
  Fiz.53,1717(1967)]}\BibitemShut {NoStop}%
\bibitem [{\citenamefont {Maki}\ \emph {et~al.}(1962)\citenamefont {Maki},
  \citenamefont {Nakagawa},\ and\ \citenamefont {Sakata}}]{Maki:1962mu}%
  \BibitemOpen
  \bibfield  {author} {\bibinfo {author} {\bibfnamefont {Z.}~\bibnamefont
  {Maki}}, \bibinfo {author} {\bibfnamefont {M.}~\bibnamefont {Nakagawa}}, \
  and\ \bibinfo {author} {\bibfnamefont {S.}~\bibnamefont {Sakata}},\ }\href
  {\doibase 10.1143/PTP.28.870} {\bibfield  {journal} {\bibinfo  {journal}
  {Prog. Theor. Phys.}\ }\textbf {\bibinfo {volume} {28}},\ \bibinfo {pages}
  {870} (\bibinfo {year} {1962})},\ \bibinfo {note} {[34(1962)]}\BibitemShut
  {NoStop}%
\bibitem [{\citenamefont {Wolfenstein}(1983)}]{Wolfenstein:1983yz}%
  \BibitemOpen
  \bibfield  {author} {\bibinfo {author} {\bibfnamefont {L.}~\bibnamefont
  {Wolfenstein}},\ }\href {\doibase 10.1103/PhysRevLett.51.1945} {\bibfield
  {journal} {\bibinfo  {journal} {Phys. Rev. Lett.}\ }\textbf {\bibinfo
  {volume} {51}},\ \bibinfo {pages} {1945} (\bibinfo {year}
  {1983})}\BibitemShut {NoStop}%
\bibitem [{\citenamefont {Altarelli}\ and\ \citenamefont
  {Feruglio}(2010)}]{Altarelli:2010gt}%
  \BibitemOpen
  \bibfield  {author} {\bibinfo {author} {\bibfnamefont {G.}~\bibnamefont
  {Altarelli}}\ and\ \bibinfo {author} {\bibfnamefont {F.}~\bibnamefont
  {Feruglio}},\ }\href {\doibase 10.1103/RevModPhys.82.2701} {\bibfield
  {journal} {\bibinfo  {journal} {Rev. Mod. Phys.}\ }\textbf {\bibinfo {volume}
  {82}},\ \bibinfo {pages} {2701} (\bibinfo {year} {2010})},\ \Eprint
  {http://arxiv.org/abs/1002.0211} {arXiv:1002.0211 [hep-ph]} \BibitemShut
  {NoStop}%
\bibitem [{\citenamefont {King}\ and\ \citenamefont
  {Luhn}(2013)}]{King:2013eh}%
  \BibitemOpen
  \bibfield  {author} {\bibinfo {author} {\bibfnamefont {S.~F.}\ \bibnamefont
  {King}}\ and\ \bibinfo {author} {\bibfnamefont {C.}~\bibnamefont {Luhn}},\
  }\href {\doibase 10.1088/0034-4885/76/5/056201} {\bibfield  {journal}
  {\bibinfo  {journal} {Rept. Prog. Phys.}\ }\textbf {\bibinfo {volume} {76}},\
  \bibinfo {pages} {056201} (\bibinfo {year} {2013})},\ \Eprint
  {http://arxiv.org/abs/1301.1340} {arXiv:1301.1340 [hep-ph]} \BibitemShut
  {NoStop}%
\bibitem [{\citenamefont {Xing}\ and\ \citenamefont
  {Zhao}(2016)}]{Xing:2015fdg}%
  \BibitemOpen
  \bibfield  {author} {\bibinfo {author} {\bibfnamefont {Z.-z.}\ \bibnamefont
  {Xing}}\ and\ \bibinfo {author} {\bibfnamefont {Z.-h.}\ \bibnamefont
  {Zhao}},\ }\href {\doibase 10.1088/0034-4885/79/7/076201} {\bibfield
  {journal} {\bibinfo  {journal} {Rept. Prog. Phys.}\ }\textbf {\bibinfo
  {volume} {79}},\ \bibinfo {pages} {076201} (\bibinfo {year} {2016})},\
  \Eprint {http://arxiv.org/abs/1512.04207} {arXiv:1512.04207 [hep-ph]}
  \BibitemShut {NoStop}%
\bibitem [{\citenamefont {Hall}\ \emph {et~al.}(2000)\citenamefont {Hall},
  \citenamefont {Murayama},\ and\ \citenamefont {Weiner}}]{Hall:1999sn}%
  \BibitemOpen
  \bibfield  {author} {\bibinfo {author} {\bibfnamefont {L.~J.}\ \bibnamefont
  {Hall}}, \bibinfo {author} {\bibfnamefont {H.}~\bibnamefont {Murayama}}, \
  and\ \bibinfo {author} {\bibfnamefont {N.}~\bibnamefont {Weiner}},\ }\href
  {\doibase 10.1103/PhysRevLett.84.2572} {\bibfield  {journal} {\bibinfo
  {journal} {Phys. Rev. Lett.}\ }\textbf {\bibinfo {volume} {84}},\ \bibinfo
  {pages} {2572} (\bibinfo {year} {2000})},\ \Eprint
  {http://arxiv.org/abs/hep-ph/9911341} {arXiv:hep-ph/9911341 [hep-ph]}
  \BibitemShut {NoStop}%
\bibitem [{\citenamefont {Haba}\ and\ \citenamefont
  {Murayama}(2001)}]{Haba:2000be}%
  \BibitemOpen
  \bibfield  {author} {\bibinfo {author} {\bibfnamefont {N.}~\bibnamefont
  {Haba}}\ and\ \bibinfo {author} {\bibfnamefont {H.}~\bibnamefont
  {Murayama}},\ }\href {\doibase 10.1103/PhysRevD.63.053010} {\bibfield
  {journal} {\bibinfo  {journal} {Phys. Rev.}\ }\textbf {\bibinfo {volume}
  {D63}},\ \bibinfo {pages} {053010} (\bibinfo {year} {2001})},\ \Eprint
  {http://arxiv.org/abs/hep-ph/0009174} {arXiv:hep-ph/0009174 [hep-ph]}
  \BibitemShut {NoStop}%
\bibitem [{\citenamefont {de~Gouvea}\ and\ \citenamefont
  {Murayama}(2003)}]{deGouvea:2003xe}%
  \BibitemOpen
  \bibfield  {author} {\bibinfo {author} {\bibfnamefont {A.}~\bibnamefont
  {de~Gouvea}}\ and\ \bibinfo {author} {\bibfnamefont {H.}~\bibnamefont
  {Murayama}},\ }\href {\doibase 10.1016/j.physletb.2003.08.045} {\bibfield
  {journal} {\bibinfo  {journal} {Phys. Lett.}\ }\textbf {\bibinfo {volume}
  {B573}},\ \bibinfo {pages} {94} (\bibinfo {year} {2003})},\ \Eprint
  {http://arxiv.org/abs/hep-ph/0301050} {arXiv:hep-ph/0301050 [hep-ph]}
  \BibitemShut {NoStop}%
\bibitem [{\citenamefont {Ge}\ \emph {et~al.}(2018)\citenamefont {Ge},
  \citenamefont {Kusenko},\ and\ \citenamefont {Yanagida}}]{Ge:2018ofp}%
  \BibitemOpen
  \bibfield  {author} {\bibinfo {author} {\bibfnamefont {S.-F.}\ \bibnamefont
  {Ge}}, \bibinfo {author} {\bibfnamefont {A.}~\bibnamefont {Kusenko}}, \ and\
  \bibinfo {author} {\bibfnamefont {T.~T.}\ \bibnamefont {Yanagida}},\ }\href
  {\doibase 10.1016/j.physletb.2018.04.040} {\bibfield  {journal} {\bibinfo
  {journal} {Phys. Lett.}\ }\textbf {\bibinfo {volume} {B781}},\ \bibinfo
  {pages} {699} (\bibinfo {year} {2018})},\ \Eprint
  {http://arxiv.org/abs/1803.03888} {arXiv:1803.03888 [hep-ph]} \BibitemShut
  {NoStop}%
\bibitem [{\citenamefont {Barrie}\ \emph {et~al.}(2020)\citenamefont {Barrie},
  \citenamefont {Ge},\ and\ \citenamefont {Yanagida}}]{Barrie:2019pqc}%
  \BibitemOpen
  \bibfield  {author} {\bibinfo {author} {\bibfnamefont {N.~D.}\ \bibnamefont
  {Barrie}}, \bibinfo {author} {\bibfnamefont {S.-F.}\ \bibnamefont {Ge}}, \
  and\ \bibinfo {author} {\bibfnamefont {T.~T.}\ \bibnamefont {Yanagida}},\
  }\href {\doibase 10.1016/j.physletb.2019.135159} {\bibfield  {journal}
  {\bibinfo  {journal} {Phys. Lett.}\ }\textbf {\bibinfo {volume} {B801}},\
  \bibinfo {pages} {135159} (\bibinfo {year} {2020})},\ \Eprint
  {http://arxiv.org/abs/1911.07430} {arXiv:1911.07430 [hep-ph]} \BibitemShut
  {NoStop}%
\bibitem [{\citenamefont {Abi}\ \emph {et~al.}(2020)\citenamefont {Abi} \emph
  {et~al.}}]{Abi:2020evt}%
  \BibitemOpen
  \bibfield  {author} {\bibinfo {author} {\bibfnamefont {B.}~\bibnamefont
  {Abi}} \emph {et~al.} (\bibinfo {collaboration} {DUNE}),\ }\href@noop {} {\
  (\bibinfo {year} {2020})},\ \Eprint {http://arxiv.org/abs/2002.03005}
  {arXiv:2002.03005 [hep-ex]} \BibitemShut {NoStop}%
\bibitem [{\citenamefont {Abe}\ \emph {et~al.}(2014)\citenamefont {Abe} \emph
  {et~al.}}]{Abe:2014oxa}%
  \BibitemOpen
  \bibfield  {author} {\bibinfo {author} {\bibfnamefont {K.}~\bibnamefont
  {Abe}} \emph {et~al.} (\bibinfo {collaboration} {Hyper-Kamiokande Working
  Group})\ }(\bibinfo {year} {2014})\ \Eprint {http://arxiv.org/abs/1412.4673}
  {arXiv:1412.4673 [physics.ins-det]} \BibitemShut {NoStop}%
\bibitem [{\citenamefont {Djurcic}\ \emph {et~al.}(2015)\citenamefont {Djurcic}
  \emph {et~al.}}]{Djurcic:2015vqa}%
  \BibitemOpen
  \bibfield  {author} {\bibinfo {author} {\bibfnamefont {Z.}~\bibnamefont
  {Djurcic}} \emph {et~al.} (\bibinfo {collaboration} {JUNO}),\ }\href@noop {}
  {\  (\bibinfo {year} {2015})},\ \Eprint {http://arxiv.org/abs/1508.07166}
  {arXiv:1508.07166 [physics.ins-det]} \BibitemShut {NoStop}%
\bibitem [{\citenamefont {Capozzi}\ \emph {et~al.}(2017)\citenamefont
  {Capozzi}, \citenamefont {Di~Valentino}, \citenamefont {Lisi}, \citenamefont
  {Marrone}, \citenamefont {Melchiorri},\ and\ \citenamefont
  {Palazzo}}]{Capozzi:2017ipn}%
  \BibitemOpen
  \bibfield  {author} {\bibinfo {author} {\bibfnamefont {F.}~\bibnamefont
  {Capozzi}}, \bibinfo {author} {\bibfnamefont {E.}~\bibnamefont
  {Di~Valentino}}, \bibinfo {author} {\bibfnamefont {E.}~\bibnamefont {Lisi}},
  \bibinfo {author} {\bibfnamefont {A.}~\bibnamefont {Marrone}}, \bibinfo
  {author} {\bibfnamefont {A.}~\bibnamefont {Melchiorri}}, \ and\ \bibinfo
  {author} {\bibfnamefont {A.}~\bibnamefont {Palazzo}},\ }\href {\doibase
  10.1103/PhysRevD.95.096014} {\bibfield  {journal} {\bibinfo  {journal} {Phys.
  Rev.}\ }\textbf {\bibinfo {volume} {D95}},\ \bibinfo {pages} {096014}
  (\bibinfo {year} {2017})},\ \Eprint {http://arxiv.org/abs/1703.04471}
  {arXiv:1703.04471 [hep-ph]} \BibitemShut {NoStop}%
\bibitem [{\citenamefont {Esteban}\ \emph {et~al.}(2019)\citenamefont
  {Esteban}, \citenamefont {Gonzalez-Garcia}, \citenamefont
  {Hernandez-Cabezudo}, \citenamefont {Maltoni},\ and\ \citenamefont
  {Schwetz}}]{Esteban:2018azc}%
  \BibitemOpen
  \bibfield  {author} {\bibinfo {author} {\bibfnamefont {I.}~\bibnamefont
  {Esteban}}, \bibinfo {author} {\bibfnamefont {M.~C.}\ \bibnamefont
  {Gonzalez-Garcia}}, \bibinfo {author} {\bibfnamefont {A.}~\bibnamefont
  {Hernandez-Cabezudo}}, \bibinfo {author} {\bibfnamefont {M.}~\bibnamefont
  {Maltoni}}, \ and\ \bibinfo {author} {\bibfnamefont {T.}~\bibnamefont
  {Schwetz}},\ }\href {\doibase 10.1007/JHEP01(2019)106} {\bibfield  {journal}
  {\bibinfo  {journal} {JHEP}\ }\textbf {\bibinfo {volume} {01}},\ \bibinfo
  {pages} {106} (\bibinfo {year} {2019})},\ \Eprint
  {http://arxiv.org/abs/1811.05487} {arXiv:1811.05487 [hep-ph]} \BibitemShut
  {NoStop}%
\bibitem [{\citenamefont {De~Salas}\ \emph {et~al.}(2018)\citenamefont
  {De~Salas}, \citenamefont {Gariazzo}, \citenamefont {Mena}, \citenamefont
  {Ternes},\ and\ \citenamefont {Tórtola}}]{deSalas:2018bym}%
  \BibitemOpen
  \bibfield  {author} {\bibinfo {author} {\bibfnamefont {P.~F.}\ \bibnamefont
  {De~Salas}}, \bibinfo {author} {\bibfnamefont {S.}~\bibnamefont {Gariazzo}},
  \bibinfo {author} {\bibfnamefont {O.}~\bibnamefont {Mena}}, \bibinfo {author}
  {\bibfnamefont {C.~A.}\ \bibnamefont {Ternes}}, \ and\ \bibinfo {author}
  {\bibfnamefont {M.}~\bibnamefont {Tórtola}},\ }\href {\doibase
  10.3389/fspas.2018.00036} {\bibfield  {journal} {\bibinfo  {journal} {Front.
  Astron. Space Sci.}\ }\textbf {\bibinfo {volume} {5}},\ \bibinfo {pages} {36}
  (\bibinfo {year} {2018})},\ \Eprint {http://arxiv.org/abs/1806.11051}
  {arXiv:1806.11051 [hep-ph]} \BibitemShut {NoStop}%
\bibitem [{\citenamefont {Capozzi}\ \emph {et~al.}(2018)\citenamefont
  {Capozzi}, \citenamefont {Lisi}, \citenamefont {Marrone},\ and\ \citenamefont
  {Palazzo}}]{Capozzi:2018ubv}%
  \BibitemOpen
  \bibfield  {author} {\bibinfo {author} {\bibfnamefont {F.}~\bibnamefont
  {Capozzi}}, \bibinfo {author} {\bibfnamefont {E.}~\bibnamefont {Lisi}},
  \bibinfo {author} {\bibfnamefont {A.}~\bibnamefont {Marrone}}, \ and\
  \bibinfo {author} {\bibfnamefont {A.}~\bibnamefont {Palazzo}},\ }\href
  {\doibase 10.1016/j.ppnp.2018.05.005} {\bibfield  {journal} {\bibinfo
  {journal} {Prog. Part. Nucl. Phys.}\ }\textbf {\bibinfo {volume} {102}},\
  \bibinfo {pages} {48} (\bibinfo {year} {2018})},\ \Eprint
  {http://arxiv.org/abs/1804.09678} {arXiv:1804.09678 [hep-ph]} \BibitemShut
  {NoStop}%
\bibitem [{\citenamefont {Kelly}\ \emph {et~al.}(2020)\citenamefont {Kelly},
  \citenamefont {Machado}, \citenamefont {Parke}, \citenamefont
  {Perez~Gonzalez},\ and\ \citenamefont {Zukanovich-Funchal}}]{Kelly:2020fkv}%
  \BibitemOpen
  \bibfield  {author} {\bibinfo {author} {\bibfnamefont {K.~J.}\ \bibnamefont
  {Kelly}}, \bibinfo {author} {\bibfnamefont {P.~A.}\ \bibnamefont {Machado}},
  \bibinfo {author} {\bibfnamefont {S.~J.}\ \bibnamefont {Parke}}, \bibinfo
  {author} {\bibfnamefont {Y.~F.}\ \bibnamefont {Perez~Gonzalez}}, \ and\
  \bibinfo {author} {\bibfnamefont {R.}~\bibnamefont {Zukanovich-Funchal}},\
  }\href@noop {} {\  (\bibinfo {year} {2020})},\ \Eprint
  {http://arxiv.org/abs/2007.08526} {arXiv:2007.08526 [hep-ph]} \BibitemShut
  {NoStop}%
\bibitem [{\citenamefont {Esteban}\ \emph {et~al.}(2020)\citenamefont
  {Esteban}, \citenamefont {Gonzalez-Garcia}, \citenamefont {Maltoni},
  \citenamefont {Schwetz},\ and\ \citenamefont {Zhou}}]{Esteban:2020cvm}%
  \BibitemOpen
  \bibfield  {author} {\bibinfo {author} {\bibfnamefont {I.}~\bibnamefont
  {Esteban}}, \bibinfo {author} {\bibfnamefont {M.}~\bibnamefont
  {Gonzalez-Garcia}}, \bibinfo {author} {\bibfnamefont {M.}~\bibnamefont
  {Maltoni}}, \bibinfo {author} {\bibfnamefont {T.}~\bibnamefont {Schwetz}}, \
  and\ \bibinfo {author} {\bibfnamefont {A.}~\bibnamefont {Zhou}},\ }\href
  {\doibase 10.1007/JHEP09(2020)178} {\bibfield  {journal} {\bibinfo  {journal}
  {JHEP}\ }\textbf {\bibinfo {volume} {09}},\ \bibinfo {pages} {178} (\bibinfo
  {year} {2020})},\ \Eprint {http://arxiv.org/abs/2007.14792} {arXiv:2007.14792
  [hep-ph]} \BibitemShut {NoStop}%
\bibitem [{\citenamefont {Ahmad}\ \emph {et~al.}(2002)\citenamefont {Ahmad}
  \emph {et~al.}}]{Ahmad:2002jz}%
  \BibitemOpen
  \bibfield  {author} {\bibinfo {author} {\bibfnamefont {Q.~R.}\ \bibnamefont
  {Ahmad}} \emph {et~al.} (\bibinfo {collaboration} {SNO}),\ }\href {\doibase
  10.1103/PhysRevLett.89.011301} {\bibfield  {journal} {\bibinfo  {journal}
  {Phys. Rev. Lett.}\ }\textbf {\bibinfo {volume} {89}},\ \bibinfo {pages}
  {011301} (\bibinfo {year} {2002})},\ \Eprint
  {http://arxiv.org/abs/nucl-ex/0204008} {arXiv:nucl-ex/0204008 [nucl-ex]}
  \BibitemShut {NoStop}%
\bibitem [{\citenamefont {An}\ \emph {et~al.}(2012)\citenamefont {An} \emph
  {et~al.}}]{An:2012eh}%
  \BibitemOpen
  \bibfield  {author} {\bibinfo {author} {\bibfnamefont {F.~P.}\ \bibnamefont
  {An}} \emph {et~al.} (\bibinfo {collaboration} {Daya Bay}),\ }\href {\doibase
  10.1103/PhysRevLett.108.171803} {\bibfield  {journal} {\bibinfo  {journal}
  {Phys. Rev. Lett.}\ }\textbf {\bibinfo {volume} {108}},\ \bibinfo {pages}
  {171803} (\bibinfo {year} {2012})},\ \Eprint {http://arxiv.org/abs/1203.1669}
  {arXiv:1203.1669 [hep-ex]} \BibitemShut {NoStop}%
\bibitem [{\citenamefont {Ahn}\ \emph {et~al.}(2012)\citenamefont {Ahn} \emph
  {et~al.}}]{Ahn:2012nd}%
  \BibitemOpen
  \bibfield  {author} {\bibinfo {author} {\bibfnamefont {J.~K.}\ \bibnamefont
  {Ahn}} \emph {et~al.} (\bibinfo {collaboration} {RENO}),\ }\href {\doibase
  10.1103/PhysRevLett.108.191802} {\bibfield  {journal} {\bibinfo  {journal}
  {Phys. Rev. Lett.}\ }\textbf {\bibinfo {volume} {108}},\ \bibinfo {pages}
  {191802} (\bibinfo {year} {2012})},\ \Eprint {http://arxiv.org/abs/1204.0626}
  {arXiv:1204.0626 [hep-ex]} \BibitemShut {NoStop}%
\bibitem [{\citenamefont {Tanabashi}\ \emph {et~al.}(2018)\citenamefont
  {Tanabashi} \emph {et~al.}}]{Tanabashi:2018oca}%
  \BibitemOpen
  \bibfield  {author} {\bibinfo {author} {\bibfnamefont {M.}~\bibnamefont
  {Tanabashi}} \emph {et~al.} (\bibinfo {collaboration} {Particle Data
  Group}),\ }\href {\doibase 10.1103/PhysRevD.98.030001} {\bibfield  {journal}
  {\bibinfo  {journal} {Phys. Rev.}\ }\textbf {\bibinfo {volume} {D98}},\
  \bibinfo {pages} {030001} (\bibinfo {year} {2018})}\BibitemShut {NoStop}%
\bibitem [{\citenamefont {Agafonova}\ \emph {et~al.}(2018)\citenamefont
  {Agafonova} \emph {et~al.}}]{Agafonova:2018auq}%
  \BibitemOpen
  \bibfield  {author} {\bibinfo {author} {\bibfnamefont {N.}~\bibnamefont
  {Agafonova}} \emph {et~al.} (\bibinfo {collaboration} {OPERA}),\ }\href
  {\doibase 10.1103/PhysRevLett.121.139901, 10.1103/PhysRevLett.120.211801}
  {\bibfield  {journal} {\bibinfo  {journal} {Phys. Rev. Lett.}\ }\textbf
  {\bibinfo {volume} {120}},\ \bibinfo {pages} {211801} (\bibinfo {year}
  {2018})},\ \bibinfo {note} {[Erratum: Phys. Rev.
  Lett.121,no.13,139901(2018)]},\ \Eprint {http://arxiv.org/abs/1804.04912}
  {arXiv:1804.04912 [hep-ex]} \BibitemShut {NoStop}%
\bibitem [{\citenamefont {Li}\ \emph {et~al.}(2018)\citenamefont {Li} \emph
  {et~al.}}]{Li:2017dbe}%
  \BibitemOpen
  \bibfield  {author} {\bibinfo {author} {\bibfnamefont {Z.}~\bibnamefont {Li}}
  \emph {et~al.} (\bibinfo {collaboration} {Super-Kamiokande}),\ }\href
  {\doibase 10.1103/PhysRevD.98.052006} {\bibfield  {journal} {\bibinfo
  {journal} {Phys. Rev.}\ }\textbf {\bibinfo {volume} {D98}},\ \bibinfo {pages}
  {052006} (\bibinfo {year} {2018})},\ \Eprint
  {http://arxiv.org/abs/1711.09436} {arXiv:1711.09436 [hep-ex]} \BibitemShut
  {NoStop}%
\bibitem [{\citenamefont {Aartsen}\ \emph {et~al.}(2019)\citenamefont {Aartsen}
  \emph {et~al.}}]{Aartsen:2019tjl}%
  \BibitemOpen
  \bibfield  {author} {\bibinfo {author} {\bibfnamefont {M.~G.}\ \bibnamefont
  {Aartsen}} \emph {et~al.} (\bibinfo {collaboration} {IceCube}),\ }\href
  {\doibase 10.1103/PhysRevD.99.032007} {\bibfield  {journal} {\bibinfo
  {journal} {Phys. Rev.}\ }\textbf {\bibinfo {volume} {D99}},\ \bibinfo {pages}
  {032007} (\bibinfo {year} {2019})},\ \Eprint
  {http://arxiv.org/abs/1901.05366} {arXiv:1901.05366 [hep-ex]} \BibitemShut
  {NoStop}%
\bibitem [{\citenamefont {Jarlskog}(1985)}]{Jarlskog:1985ht}%
  \BibitemOpen
  \bibfield  {author} {\bibinfo {author} {\bibfnamefont {C.}~\bibnamefont
  {Jarlskog}},\ }\href {\doibase 10.1103/PhysRevLett.55.1039} {\bibfield
  {journal} {\bibinfo  {journal} {Phys. Rev. Lett.}\ }\textbf {\bibinfo
  {volume} {55}},\ \bibinfo {pages} {1039} (\bibinfo {year}
  {1985})}\BibitemShut {NoStop}%
\bibitem [{\citenamefont {Denton}\ \emph {et~al.}(2016)\citenamefont {Denton},
  \citenamefont {Minakata},\ and\ \citenamefont {Parke}}]{Denton:2016wmg}%
  \BibitemOpen
  \bibfield  {author} {\bibinfo {author} {\bibfnamefont {P.~B.}\ \bibnamefont
  {Denton}}, \bibinfo {author} {\bibfnamefont {H.}~\bibnamefont {Minakata}}, \
  and\ \bibinfo {author} {\bibfnamefont {S.~J.}\ \bibnamefont {Parke}},\ }\href
  {\doibase 10.1007/JHEP06(2016)051} {\bibfield  {journal} {\bibinfo  {journal}
  {JHEP}\ }\textbf {\bibinfo {volume} {06}},\ \bibinfo {pages} {051} (\bibinfo
  {year} {2016})},\ \Eprint {http://arxiv.org/abs/1604.08167} {arXiv:1604.08167
  [hep-ph]} \BibitemShut {NoStop}%
\bibitem [{\citenamefont {Denton}(2020)}]{peter_b_denton_2020_3709501}%
  \BibitemOpen
  \bibfield  {author} {\bibinfo {author} {\bibfnamefont {P.~B.}\ \bibnamefont
  {Denton}},\ }\href {\doibase 10.5281/zenodo.3709501} {\enquote {\bibinfo
  {title} {{PeterDenton/Neutrino-Normalcy v1.0.0}},}\ }\bibinfo {howpublished}
  {\url{https://github.com/PeterDenton/Neutrino-Normalcy}} (\bibinfo {year}
  {2020})\BibitemShut {NoStop}%
\bibitem [{\citenamefont {Parke}\ and\ \citenamefont
  {Ross-Lonergan}(2016)}]{Parke:2015goa}%
  \BibitemOpen
  \bibfield  {author} {\bibinfo {author} {\bibfnamefont {S.}~\bibnamefont
  {Parke}}\ and\ \bibinfo {author} {\bibfnamefont {M.}~\bibnamefont
  {Ross-Lonergan}},\ }\href {\doibase 10.1103/PhysRevD.93.113009} {\bibfield
  {journal} {\bibinfo  {journal} {Phys. Rev.}\ }\textbf {\bibinfo {volume}
  {D93}},\ \bibinfo {pages} {113009} (\bibinfo {year} {2016})},\ \Eprint
  {http://arxiv.org/abs/1508.05095} {arXiv:1508.05095 [hep-ph]} \BibitemShut
  {NoStop}%
\bibitem [{\citenamefont {Vissani}(1997)}]{Vissani:1997pa}%
  \BibitemOpen
  \bibfield  {author} {\bibinfo {author} {\bibfnamefont {F.}~\bibnamefont
  {Vissani}},\ }\href@noop {} {\  (\bibinfo {year} {1997})},\ \Eprint
  {http://arxiv.org/abs/hep-ph/9708483} {arXiv:hep-ph/9708483} \BibitemShut
  {NoStop}%
\bibitem [{\citenamefont {Harrison}\ \emph {et~al.}(1995)\citenamefont
  {Harrison}, \citenamefont {Perkins},\ and\ \citenamefont
  {Scott}}]{Harrison:1994iv}%
  \BibitemOpen
  \bibfield  {author} {\bibinfo {author} {\bibfnamefont {P.}~\bibnamefont
  {Harrison}}, \bibinfo {author} {\bibfnamefont {D.}~\bibnamefont {Perkins}}, \
  and\ \bibinfo {author} {\bibfnamefont {W.}~\bibnamefont {Scott}},\ }\href
  {\doibase 10.1016/0370-2693(95)00213-5} {\bibfield  {journal} {\bibinfo
  {journal} {Phys. Lett. B}\ }\textbf {\bibinfo {volume} {349}},\ \bibinfo
  {pages} {137} (\bibinfo {year} {1995})}\BibitemShut {NoStop}%
\bibitem [{\citenamefont {Harrison}\ \emph {et~al.}(2002)\citenamefont
  {Harrison}, \citenamefont {Perkins},\ and\ \citenamefont
  {Scott}}]{Harrison:2002er}%
  \BibitemOpen
  \bibfield  {author} {\bibinfo {author} {\bibfnamefont {P.~F.}\ \bibnamefont
  {Harrison}}, \bibinfo {author} {\bibfnamefont {D.~H.}\ \bibnamefont
  {Perkins}}, \ and\ \bibinfo {author} {\bibfnamefont {W.~G.}\ \bibnamefont
  {Scott}},\ }\href {\doibase 10.1016/S0370-2693(02)01336-9} {\bibfield
  {journal} {\bibinfo  {journal} {Phys. Lett.}\ }\textbf {\bibinfo {volume}
  {B530}},\ \bibinfo {pages} {167} (\bibinfo {year} {2002})},\ \Eprint
  {http://arxiv.org/abs/hep-ph/0202074} {arXiv:hep-ph/0202074 [hep-ph]}
  \BibitemShut {NoStop}%
\bibitem [{\citenamefont {Everett}\ and\ \citenamefont
  {Stuart}(2009)}]{Everett:2008et}%
  \BibitemOpen
  \bibfield  {author} {\bibinfo {author} {\bibfnamefont {L.~L.}\ \bibnamefont
  {Everett}}\ and\ \bibinfo {author} {\bibfnamefont {A.~J.}\ \bibnamefont
  {Stuart}},\ }\href {\doibase 10.1103/PhysRevD.79.085005} {\bibfield
  {journal} {\bibinfo  {journal} {Phys. Rev. D}\ }\textbf {\bibinfo {volume}
  {79}},\ \bibinfo {pages} {085005} (\bibinfo {year} {2009})},\ \Eprint
  {http://arxiv.org/abs/0812.1057} {arXiv:0812.1057 [hep-ph]} \BibitemShut
  {NoStop}%
\bibitem [{\citenamefont {Chen}\ \emph {et~al.}(2018)\citenamefont {Chen},
  \citenamefont {Centelles~Chuli\'a}, \citenamefont {Ding}, \citenamefont
  {Srivastava},\ and\ \citenamefont {Valle}}]{Chen:2018eou}%
  \BibitemOpen
  \bibfield  {author} {\bibinfo {author} {\bibfnamefont {P.}~\bibnamefont
  {Chen}}, \bibinfo {author} {\bibfnamefont {S.}~\bibnamefont
  {Centelles~Chuli\'a}}, \bibinfo {author} {\bibfnamefont {G.-J.}\ \bibnamefont
  {Ding}}, \bibinfo {author} {\bibfnamefont {R.}~\bibnamefont {Srivastava}}, \
  and\ \bibinfo {author} {\bibfnamefont {J.~W.}\ \bibnamefont {Valle}},\ }\href
  {\doibase 10.1103/PhysRevD.98.055019} {\bibfield  {journal} {\bibinfo
  {journal} {Phys. Rev. D}\ }\textbf {\bibinfo {volume} {98}},\ \bibinfo
  {pages} {055019} (\bibinfo {year} {2018})},\ \Eprint
  {http://arxiv.org/abs/1806.03367} {arXiv:1806.03367 [hep-ph]} \BibitemShut
  {NoStop}%
\bibitem [{\citenamefont {Abe}\ \emph {et~al.}(2018)\citenamefont {Abe} \emph
  {et~al.}}]{Abe:2018wpn}%
  \BibitemOpen
  \bibfield  {author} {\bibinfo {author} {\bibfnamefont {K.}~\bibnamefont
  {Abe}} \emph {et~al.} (\bibinfo {collaboration} {T2K}),\ }\href {\doibase
  10.1103/PhysRevLett.121.171802} {\bibfield  {journal} {\bibinfo  {journal}
  {Phys. Rev. Lett.}\ }\textbf {\bibinfo {volume} {121}},\ \bibinfo {pages}
  {171802} (\bibinfo {year} {2018})},\ \Eprint
  {http://arxiv.org/abs/1807.07891} {arXiv:1807.07891 [hep-ex]} \BibitemShut
  {NoStop}%
\end{thebibliography}%

\end{document}